\documentclass[conference]{IEEEtran}

\usepackage{blindtext, graphicx}

\usepackage{url}

\usepackage{xargs}                      
\usepackage[pdftex,dvipsnames]{xcolor}  
\usepackage[colorinlistoftodos,prependcaption,textsize=medium]{todonotes}
\newcommandx{\unsure}[2][1=]{\todo[linecolor=red,backgroundcolor=red!25,bordercolor=red,#1]{#2}}
\newcommandx{\change}[2][1=]{\todo[linecolor=blue,backgroundcolor=blue!25,bordercolor=blue,#1]{#2}}
\newcommandx{\info}[2][1=]{\todo[inline, linecolor=OliveGreen,backgroundcolor=OliveGreen!25,bordercolor=OliveGreen,#1]{#2}}
\newcommandx{\improvement}[2][1=]{\todo[linecolor=Plum,backgroundcolor=Plum!25,bordercolor=Plum,#1]{#2}}
\newcommandx{\thiswillnotshow}[2][1=]{\todo[disable,#1]{#2}}
\ifCLASSINFOpdf
\else
\fi
\begin{document}

%
\title{Reservoir Computing Using \\ Non-Uniform Binary Cellular Automata}


\author{
\IEEEauthorblockN{Stefano Nichele}
\IEEEauthorblockA{Department of Computer Science\\
Oslo and Akershus University College of Applied Sciences\\
Oslo, Norway\\
\texttt{stefano.nichele@hioa.no}}
\and
\IEEEauthorblockN{Magnus S. Gundersen}
\IEEEauthorblockA{Department of Computer and Information Science\\
Norwegian University of Science and Technology\\
Trondheim, Norway\\
\texttt{magnugun@stud.ntnu.no}}

}
 

%


\maketitle

\begin{abstract}
The Reservoir Computing (RC) paradigm utilizes a dynamical system, i.e., a reservoir, and a linear classifier, i.e., a read-out layer, to process data from sequential classification tasks. In this paper the usage of Cellular Automata (CA) as a reservoir is investigated. The use of CA in RC has been showing promising results. In this paper, selected state-of-the-art experiments are reproduced. It is shown that some CA-rules perform better than others, and the reservoir performance is improved by increasing the size of the CA reservoir itself.  
In addition, the usage of parallel loosely coupled CA-reservoirs, where each reservoir has a different CA-rule, is investigated. The experiments performed on quasi-uniform CA reservoir provide valuable insights in CA-reservoir design. The results herein show that some rules do not work well together, while other combinations work remarkably well. This suggests that non-uniform CA could represent a powerful tool for novel CA reservoir implementations.

\end{abstract}

\begin{IEEEkeywords}
Reservoir Computing, Cellular Automata, Parallel Reservoir, Recurrent Neural Networks, Non-Uniform Cellular Automata.
\end{IEEEkeywords}

%
\IEEEpeerreviewmaketitle

\section{Introduction}
Real life problems often require processing of time-series data. Systems that process such data must remember inputs from previous time-steps in order to make correct predictions in future time-step, i.e, they must have some sort of memory. Recurrent Neural Networks (RNN) have been shown to possess such memory \cite{Goodfellow-et-al-2016-Book}.

Unfortunately, training RNNs using traditional methods, i.e., gradient descent, is difficult \cite{bengio1994learning}. A fairly novel approach called Reservoir Computing (RC) has been proposed \cite{jaeger2001echo, natschlager2002liquid} to mitigate this problem. RC splits the RNN into two parts; the non-trained recurrent part, i.e., a reservoir, and the trainable feed-forward part, i.e. a read-out layer.

In this paper, an RC-system is investigated, and a computational model called Cellular Automata (CA) \cite{von1966theory} is used as the reservoir. This approach to RC was proposed in \cite{yilmaz2014reservoir}, and further studied in \cite{yilmaz2015connectionist}, \cite{bye2016master}, and \cite{margem2016experimental}. The term ReCA is used as an abbreviation for "Reservoir Computing using Cellular Automata", and is adopted from the latter paper. 

In this paper a fully functional ReCA system is implemented and extended into a parallel CA reservoir system (loosely coupled). Various configurations of parallel reservoir are tested, and compared to the results of a single-reservoir system. This approach is discussed and insights of different configurations of CA-reservoirs are given.

\section{Background}
\subsection{Reservoir Computing} \label{background.rc}
Feed-forward Neural Networks (NNs) are neural network models without feedback-connections, i.e. they are not aware of their own outputs \cite{Goodfellow-et-al-2016-Book}. They have gained popularity because of their ability to be trained to solve classification tasks. Examples include image classification \cite{szegedy2015going}, or playing the board game GO \cite{silver2016mastering}. However, when trying to solve problems that include sequential data, such as sentence-analysis, they often fall short \cite{Goodfellow-et-al-2016-Book}. For example, sentences may have different lengths, and the important parts may be spatially separated even for sentences with equal semantics. Recurrent Neural Networks (RNNs) can overcome this problem \cite{Goodfellow-et-al-2016-Book}, being able to process sequential data through memory of previous inputs which are remembered by the network. This is done by relieving the neural network of the constraint of not having feedback-connections. However, networks with recurrent connections are notoriously difficult to train by using traditional methods \cite{bengio1994learning}.

\begin{figure}
\centering
    \includegraphics[width=2.6in]{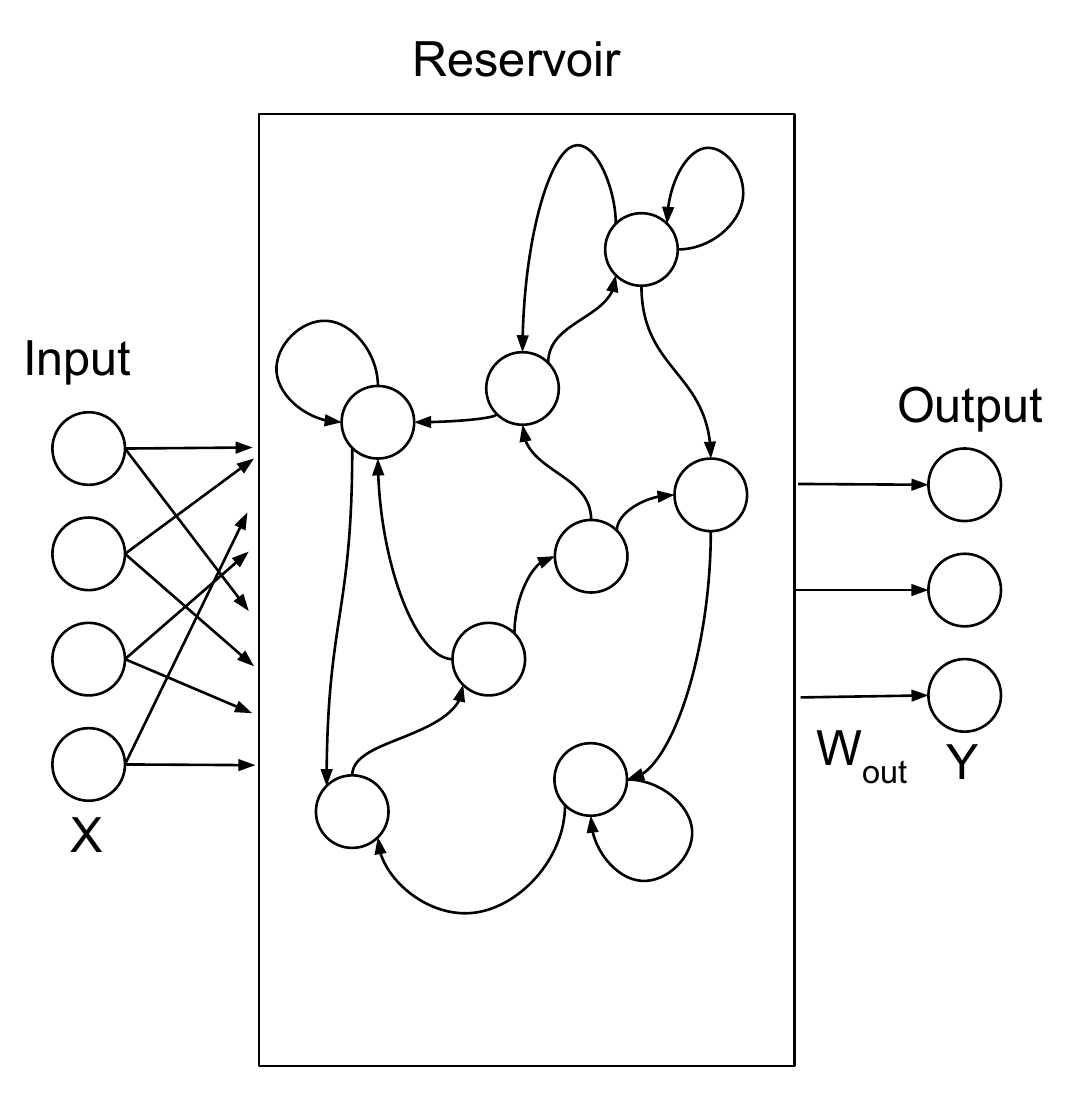}
\caption{General RC framework. Input $X$ is connected to some or all of the reservoir nodes. Output $Y$ is usually fully connected to the reservoir nodes. Only the output-weights $W_{out}$ are trained.}
\label{fig:grc}
\end{figure}

Reservoir Computing (RC) is a paradigm in machine learning that combines the powerful dynamics of an RNN with the trainability of a feed-forward neural network. The first part of an RC-system consists of an untrained RNN, called reservoir. This reservoir is connected to a trained feed-forward neural network, called readout-layer. This setup can be seen in fig. \ref{fig:grc}

The field of RC has been proposed independently by two approaches, namely Echo State Networks (ESN) \cite{jaeger2001echo} and Liquid State Machines (LSM) \cite{natschlager2002liquid}. By examining these approaches, important properties of reservoirs are outlined. 

Perhaps the most important feature is the Echo state property \cite{jaeger2001echo}. Previous inputs "echo" through the reservoir for a given number of time steps after the input has occurred, and thereby slowly disappearing without being amplified. This property is achieved in traditional RC-approaches by clever reservoir design. In the case of ESN, this is achieved by scaling of the connection weights of the recurrent nodes in the reservoir \cite{lukovsevivcius2012reservoir}.

As discussed in \cite{bertschinger2004rnnedge}, the reservoir should preferably exhibit edge of chaos behaviors \cite{langton1990computation}, in order to allow for high computational power \cite{gibbons2010unifying}.

\subsection{Various RC-approaches} \label{background.diffImpl}
Different RC-approaches use reservoir substrates that exhibit the desired properties. In \cite{fernando2003pattern} an actual bucket of water is implemented as a reservoir for speech-recognition, and in \cite{jones2007there} the E.coli-bacteria is used as a reservoir.
In \cite{snyder2013computational} and more recently in \cite{burkow2016master}, the usage of Random Boolean Networks (RBN) reservoirs is explored. RBNs can be considered as an abstraction of CA \cite{gershenson2004introduction}, and is thereby a related approach to the one presented in this paper.  

\subsection{Cellular Automata} \label{background.ca}
Cellular Automaton (CA) is a computational model, first proposed by Ulam and von Neumann in the 1940s \cite{von1966theory}. It is a complex, decentralized and highly parallel system, in which computations may emerge \cite{sipper1999emergence} through local interactions and without any form of centralized control. Some CA have been proved to be Turing complete \cite{cook2004universality}, i.e. having all properties required for computation; that is transmission, storage and modification of information \cite{langton1990computation}. 

A CA usually consists of a grid of cells, each cell with a current state. The state of a cell is determined by the update-function $f$, which is a function of the neighboring states $n$. This update-function is applied to the CA for a given number of iterations. These neighbors are defined as a number of cells in the immediate vicinity of the cell itself.

In this paper, only one-dimensional elementary CA is used. This means that the CA only consists of a one-dimensional vector of cells, named $A$, each cell with state ${S \in \{0,1\}}$. In all the figures in this paper, $S=0$ is shown as white, while $S=1$ is shown as black. The cells have three neighbors; the cell to the left, itself, and the cell to the right. A cell is a neighbor of itself by convention. The boundary conditions at each end of the 1D-vector is usually solved by wrap-around, where the leftmost cell becomes a neighbor of the rightmost, and vice versa. 

\begin{figure}
\centering
    \includegraphics[width=3.5in]{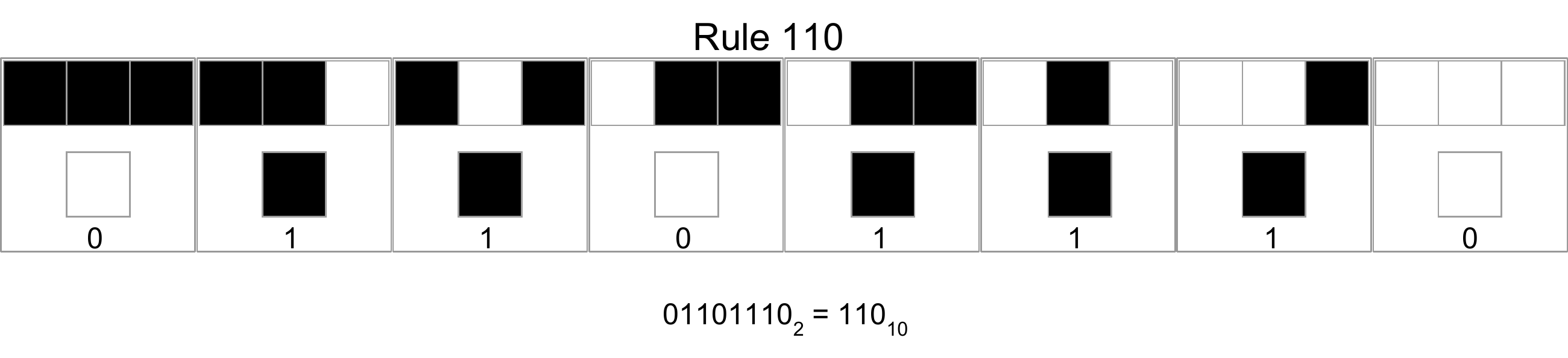}
\caption{Elementary CA rule 110. The figure depicts all the possible combinations that the neighbors of a cell can have. A cell is its own neighbor by convention.}
\label{fig:rule110}
\end{figure}

The update-function $f$, hereafter denoted rule $Z$, works accordingly by taking three binary inputs, and outputting one binary value. This results in $2^8 = 256$ different rules. An example of such a rule is shown in fig. \ref{fig:rule110}, where rule 110 is depicted. The numbering of the rules follows the naming convention described by Wolfram \cite{wolfram2002new}, where the resulting binary string is converted to a base 10 number. The CA is usually updated in synchronous steps, where all the cells in the 1D-vector are updated at the same time. One update is called an iteration, and the total number of iterations is denoted by $I$.

The rules may be divided into four qualitative classes \cite{wolfram2002new}, that exhibit different properties when evolved; class I: evolves to a static state, class II: evolves to a periodic structure, class III: evolves to chaotic patterns and class IV: evolves to complex patterns. Class I and II rules will fall into an attractor after a short while \cite{langton1990computation}, and behave orderly. Class III rules are chaotic, which means that the organization quickly descends into randomness. Class IV rules are the most interesting ones, as they reside at a phase transition between the chaotic and ordered phase, i.e., at the edge of chaos \cite{langton1990computation}. In uniform CA, all cells share the same rule, whether non-uniform CA cells are governed by different rules. Quasi-uniform CA are non-uniform with a small number of diverse rules.

\subsection{Cellular automata in reservoir computing} \label{background.rcca}
\begin{figure}
\centering
    \includegraphics[width=2.6in]{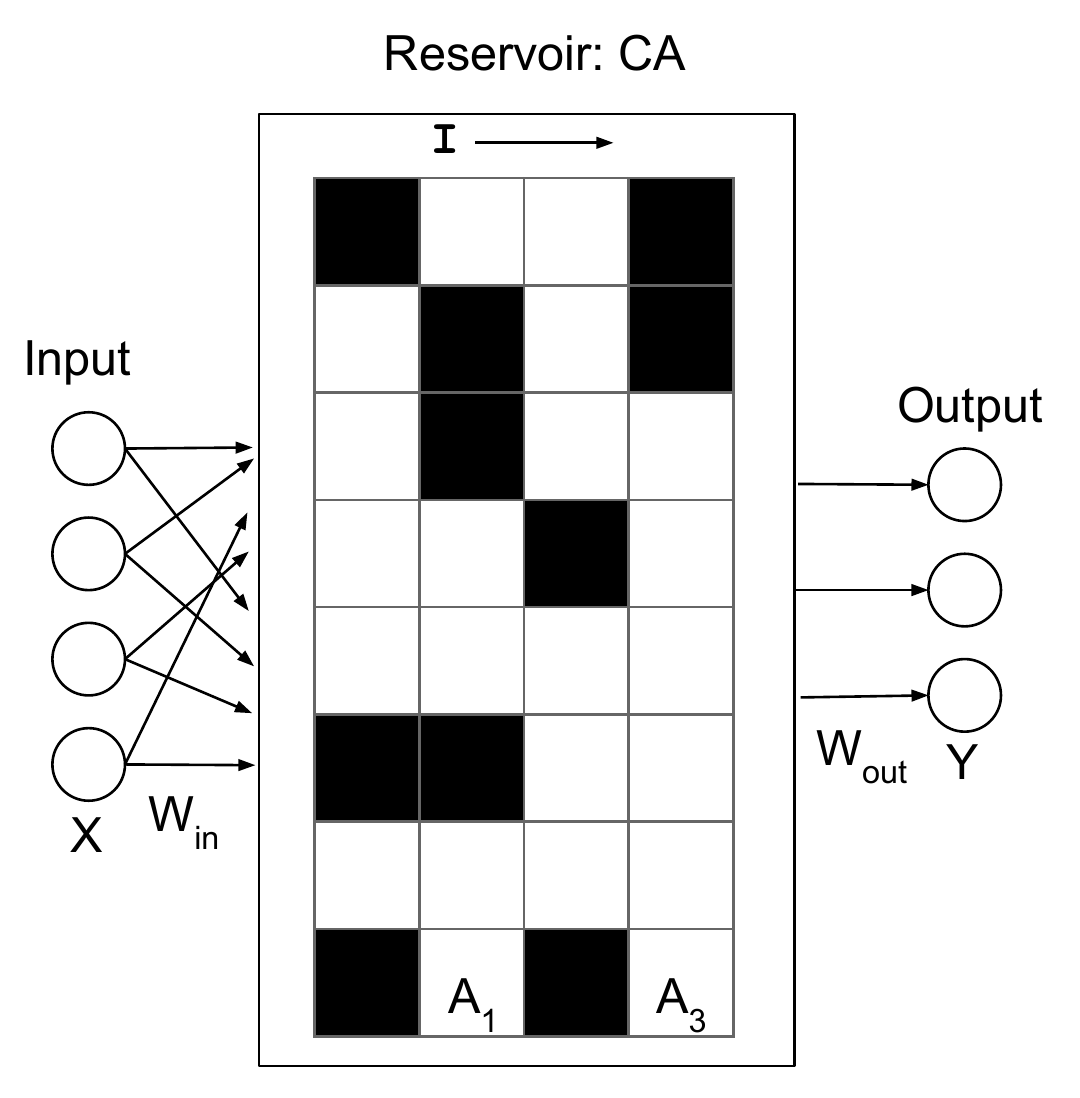}
\caption{General ReCA framework. Input $X$ is projected onto the cells of a one dimensional (1D) cellular automata, and the CA-rule is applied for a number $I$ of iterations. In the figure, each iteration is stored, and denoted by $A_i$. The readout-layer weights $W_{out}$ are trained according to the target-function. Figure adapted from \cite{yilmaz2015connectionist}}
\label{fig:grcca}
\end{figure}

As proposed in \cite{yilmaz2014reservoir}, CA may be used as reservoir of dynamical systems. The conceptual overview is shown in fig. \ref{fig:grcca}. Such system is referred to as ReCA in \cite{margem2016experimental}, and the same name is therefore adopted in this paper. The projection of the input to the CA-reservoir can be done in two different ways \cite{yilmaz2014reservoir}. If the input is binary, the projection is straightforward, where each feature dimension of the input is mapped to a cell. If the input is non-binary, the projection can be done by a weighted summation from the input to each cell. See \cite{yilmaz2015connectionist} for more details. 

The time-evolution of the reservoir can be represented as follows:
\[ A_1 = Z(A_0)\]
\[ A_2 = Z(A_1)\]
\[...\]
\[ A_I = Z(A_{I-1})\]
Where ${A_m}$ is the state of the 1D CA at iteration m and Z is the CA-rule that was applied. ${A_0}$ is the initial state of the CA, often an external input, as discussed later.

As discussed in section \ref{background.rc}, a reservoir often operates at the edge of chaos \cite{gibbons2010unifying}. Selecting CA-based reservoirs that exhibit this property is trivial, as rules that lie inside Wolfram class IV can provide this property. Additionally, to fully exploit such property, all $I$ iterations of a the CA evolution are used for classification, and this can be stated as follows:
\[ {{A} = {[{A_1};{A_2};... {A_I}]}} \]
Where $A$ is used for classification. 

The ReCA system must also exhibit the echo state property, as described in section \ref{background.rc}. This is done by allowing the CA to take external input, while still remembering the current state. As descibed in more details later, ReCA-systems address this issue by using some time-transition function, named F, which allows some previous inputs to echo through the CA. 

CA also provide additional advantages to RC. In \cite{yilmaz2015connectionist} a speedup of 1.5-3X in the number of operations compared to the ESN \cite{jaeger2012long} approach is reported. This is mainly due to a CA relying on bit-wise operations, while ESN uses floating point operations. This can be additionally exploited by utilizing custom made hardware like FPGAs. In addition, if edge-of-chaos rules are selected, Turing complete computational power is present in the reservoir. CA theoretical analysis is easier than RNNs, and they allow Boolean logic and Galois field algebra.

\subsection{ReCA system implementations}
ReCA systems are a very novel concept and therefore there are only few implemented examples at the current stage of research. Yilmaz \cite{yilmaz2014reservoir, yilmaz2015connectionist} has implemented a ReCA system with elementary CA and Game of Life \cite{conway1970game}. Bye \cite{bye2016master} also demonstrated a functioning ReCA-system in his master's thesis (supervised by Nichele). The used approaches are similar, however, there are some key differences: 

\subsubsection{Encoding and random mappings} \label{background.encoding}
\begin{figure}
\centering
    \includegraphics[width=2in]{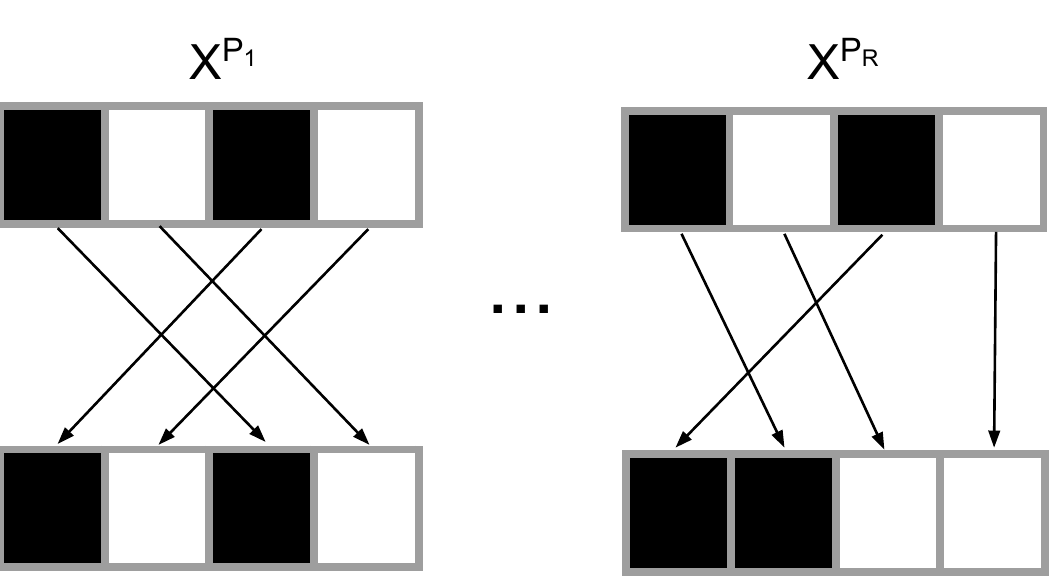}
\caption{The encoding used in \cite{yilmaz2015connectionist}. For a total of R permutations, $X$ is randomly mapped to vectors of the same size as the input-vector itself. }
\label{fig:encoding_yilmaz}
\end{figure}

In the encoding stage, \cite{yilmaz2015connectionist} used random permutations over the same input-vector. This encoding scheme can be seen in fig. \ref{fig:encoding_yilmaz}. The permutation procedure is repeated  $R$ number of times, because it was experimentally observed that multiple random mappings improve performance. 

\begin{figure}
\centering
    \includegraphics[width=3in]{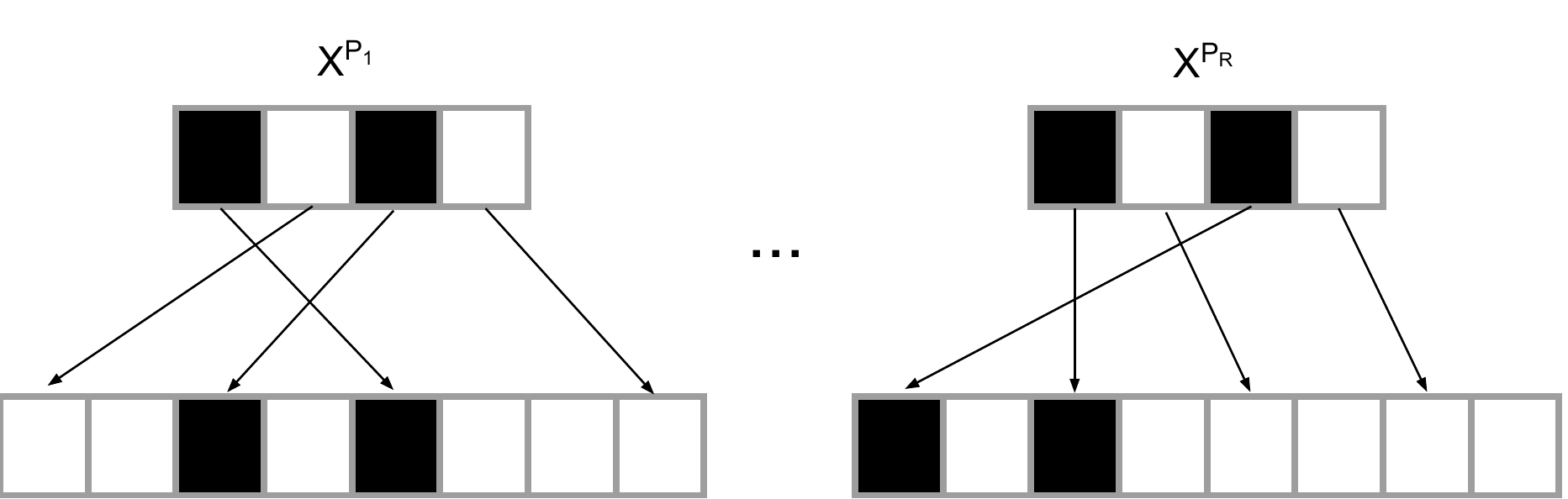}
\caption{The encoding used in \cite{bye2016master}. The input X is randomly mapped to a vector with size larger than the input vector itself. This mapping is done R times. The size of the vector that the input is mapped to can be determined in two ways. Either by "automaton size", which explicitly gives the size of the vector (in this case 8), or by the C-parameter, which the size is given by $C*|X^{p_n}|$ (in this case C=2)}
\label{fig:encoding_bye}
\end{figure}

In \cite{bye2016master} a similar approach was used. The main difference is that the input is mapped to a vector that is larger than the input-vector itself. The size of this mapping-vector is given by a parameter "automaton size". This approach can be seen in fig. \ref{fig:encoding_bye}. The input-bits are randomly mapped to one of the bits in the mapping-vector. The ones that do not have any mapping to them are left to zero.

In the work herein, the approach described in \cite{bye2016master} is used, but with a modification. Instead of using the automaton size-parameter, the C-parameter is introduced. The total length of the permutation is given by the number C multiplied by the length of the input-vector. In the case of fig. \ref{fig:encoding_bye}, the automation size would be 8, and C would be 2. 

\subsubsection{Feed-forward or recurrent} \label{background.feedfrecurrent}
\cite{yilmaz2015connectionist} proposed both a feed-forward and a recurrent design. The difference was whether the whole input-sequence is presented to the system in one chunk or step-by-step. \cite{bye2016master} only described a recurrent design. Only the recurrent architectures will be investigated in this paper. This is because it is more in line with traditional RNNs and RC-systems, and is conceptually more biologically plausible. 

\subsubsection{Concatenation of the encoded inputs before propagating into the reservoir} \label{background.concat}
After random mappings have been created, there is another difference in the proposed approaches. In the recurrent architecture, \cite{yilmaz2015connectionist} concatenates the $R$ number of permutations into one large vector of length (${R*input\_length}$) before propagating it in a reservoir of the same width as this vector. The 1D input-vector at time-step $t$ can be expressed as follows:
\[ {{X_t^P} = {[ X_t^{P_1}; X_t^{P_2}; X_t^{P_3}; ... X_t^{P_R} ]}} \]

${X_t^P}$ is inserted into the reservoir as described in section \ref{background.rcca}, and then iterated $I$ times. The iterations are then concatenated into the vector ${A^t}$, which is used for classification at time-step t. 

\[ {{A^t} = {[{A_1};{A_2};... {A_I}]}} \] \label{eq:pyls}

\cite{bye2016master} adapted a different approach, the same one that was also used by the feed-forward architecture in \cite{yilmaz2015connectionist}, where the R different permutations are iterated in separate reservoirs, and the different reservoirs are then concatenated before they are used by the classifier. The vector which is used for classification at time-step $t$ is as follows:
\[ {{A^t} = {[{A^{t}_{P_1}};{A^{t}_{P_2}};... {A^{t}_{P_R}}]}} \]

Where ${A^{t}_{P_n}}$ is the vector from the concatenated reservoir. 
In this paper, the recurrent architecture approach is used.

\subsubsection{Time-transition} \label{background.timetrans}
\begin{figure}
\centering
    \includegraphics[width=1.8in]{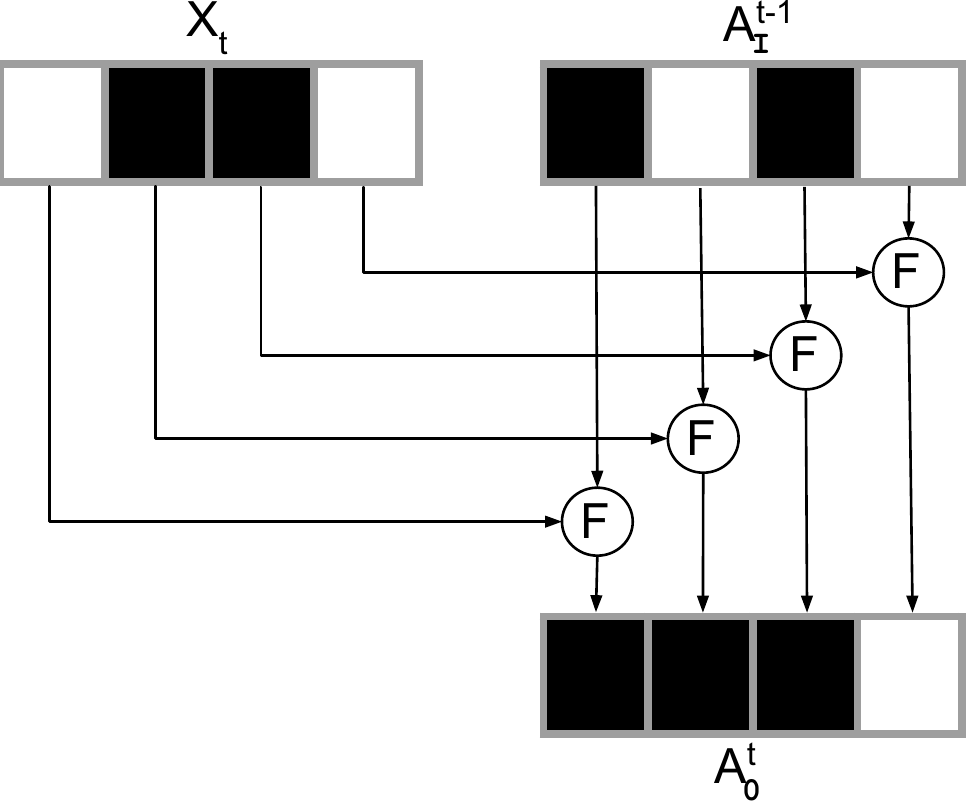}
\caption{Time transition used in \cite{yilmaz2015connectionist}. The sequence input $X_t$ is combined with the state of the reservoir at the last iteration at the previous time-step $A^{t-1}_I$. The function $F$ may be any bit-wise function. Only one permutation is shown in the figure to increase readability. }
\label{fig:yilmazTT}
\end{figure}

In order to allow the system to remember previous inputs, a time-transition function is needed to translate between the current time-step and the next. One possibility is to use normalized addition as time-transition function, as shown in fig. \ref{fig:yilmazTT}, with F as normalized addition. This function works as follows: The cell values are added, and if the sum is 2 (1+1) the output-value becomes 1, if the sum is 0, the output-value becomes 0 and if the sum is 1, the cell-value is decided randomly (0 or 1). The initial 1D-CA-vector of the reservoir at time-step $t$ is then expressed as:

\[ {{A_0} = {F(X_t, {A_I}^{t-1})},  \qquad {t>0}} \]

Where $F$ may be any bit-wise operation, $X_t$ is the input from the sequential task at time-step $t$, and ${A_I}^{t-1}$ is the last iteration of the previous time-step. At the first time-step (t=0), the transition-function is bypassed, and the input $X_t$ is used directly in the reservoir. 

\begin{figure}
\centering
    \includegraphics[width=3in]{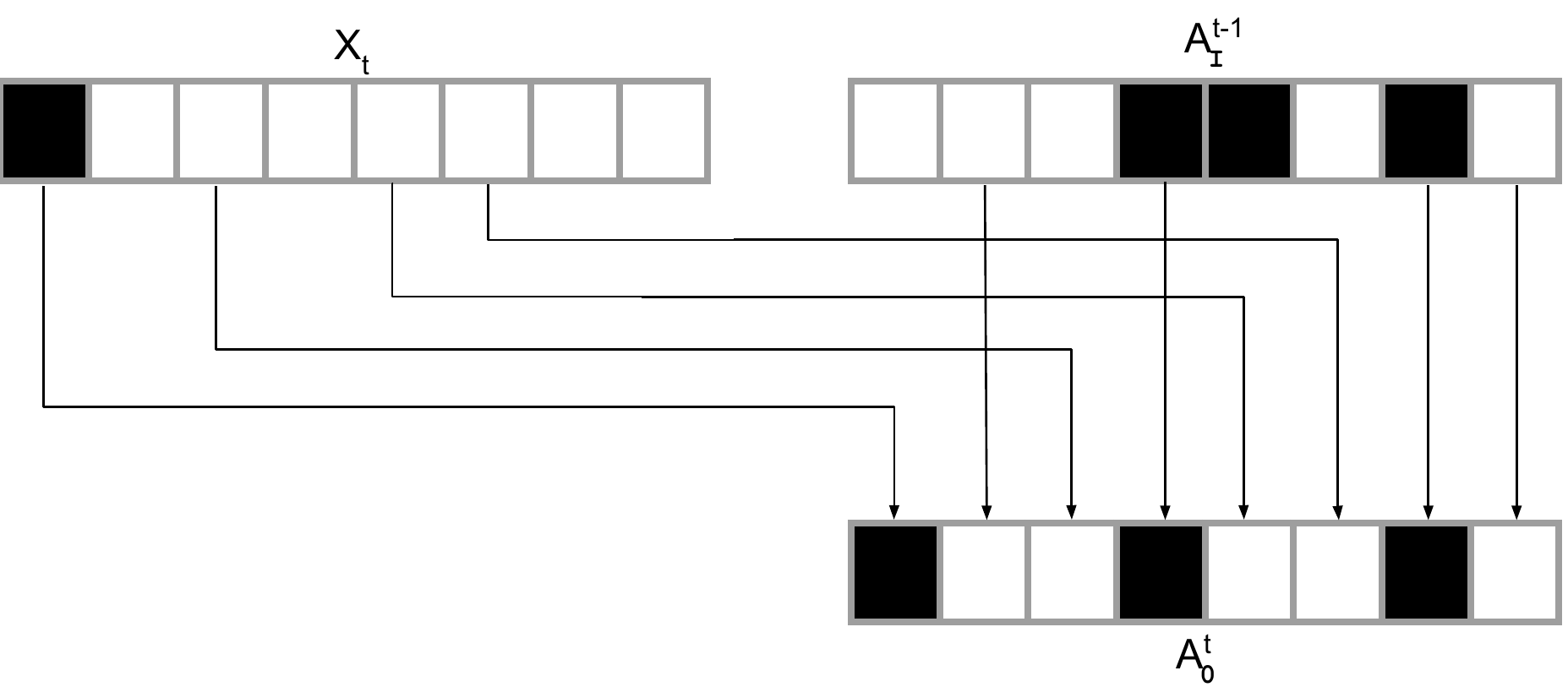}
\caption{Time-transition by permutation. The input is directly copied from $X_t$, according to the mapping from the encoder, as shown in fig. \ref{fig:encoding_bye}. The other cells have their values copied from the last iteration of the previous time-step $A^{t-1}_I$. Only one permutation is shown to increase readability. }
\label{fig:byeTT}
\end{figure}

Another possibility is to use "permutation transition" as time-transition function, as seen in fig. \ref{fig:byeTT}. Here, all cells that have a mapping to them (from the encoder) are bit-wise filled with the value of input-vector ${X}$. If the cells do not have any mapping to them, the values from ${{A_I}^{t-1}}$ are inserted. This allows the CA to have memory across time-steps in sequential tasks. By adjusting the automaton-size, or C-parameter, the interaction between each time-step can be regulated. 

The described approaches have different effects on the parameters R and I, and also the resulting size of the reservoir. This is relevant when discussing the computational complexity of  ReCA systems. 

In this paper, the "permutation transition" is used.

\section{Experimental Setup}

\begin{figure*}
\centering
    \includegraphics[width=6in]{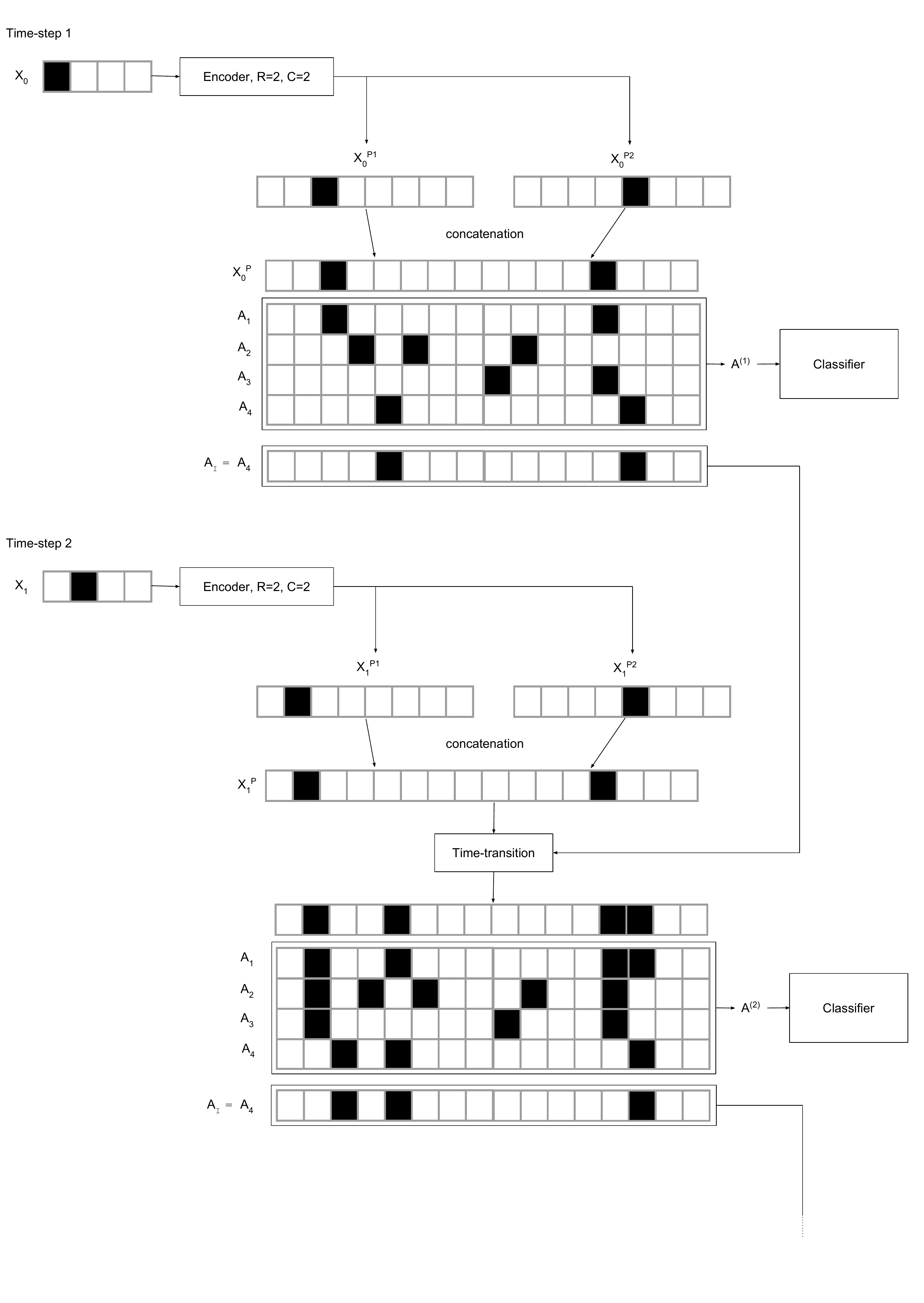}
    
\caption{Architecture of the implemented system. The encoding is done according to the encoding-scheme as shown in fig. \ref{fig:encoding_bye}, but with the slight modification of the C-parameter. The encoding is exemplified with R=2 and C=2, which yields a size of eight for each permutation. The two permutations are then concatenated. At time-step 1, there are no previous inputs, and the concatenated vector is simply used as the first iteration of the CA-reservoir. The rule Z is then applied for I iterations. At time-step 2, the encoding and concatenation is repeated. The time-transition scheme is then applied, as described in fig. \ref{fig:byeTT}. The procedure as described in time-step 2 is repeated until the end of the sequence. }
\label{fig:fullArch}
\end{figure*}

The basic architecture implemented in this paper is shown in fig. \ref{fig:fullArch}. The encoder is based on the architecture described in \cite{bye2016master}. In this paper, the parameter C is introduced as a metric on how large resulting mapping-vector should be.
The concatenation procedure is adapted from \cite{yilmaz2015connectionist}. The vectors, after the encoding (random mappings), are concatenated into one large vector. This vector is then propagated into the reservoir, as described in section \ref{background.concat}.
The time-transition function is adapted from \cite{bye2016master}. The mappings from the encoder are saved, and used as a basis where new inputs are mapped to, as described in section \ref{background.timetrans}. The values from the last step in the previous time-step are directly copied. 
The classifier used in this paper is a Support Vector Machine, as implemented in the Python machine learning framework scikit-learn \cite{pedregosa2011scikit}. 
The code-base that was used in this paper is available for download \cite{recaCode}.

\begin{figure}
\centering
    \includegraphics[height=3in]{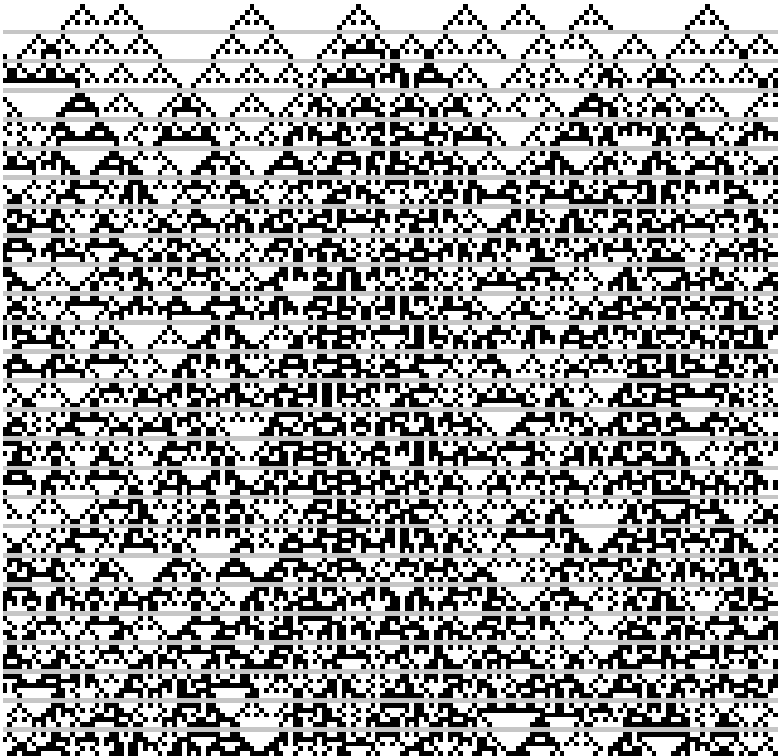}
\caption{Example run of the ReCA system with rule 90. The run is done with the parameters R=8, I=4 and C=5. The horizontal gray lines represent a time-step, in which the time-transition function is applied to every bit. Time flows downwards. The visualization is produced with the ReCA system described in this paper. }
\label{fig:rcca_single}
\end{figure}

\begin{figure}
\centering
    \includegraphics[width=2.2in]{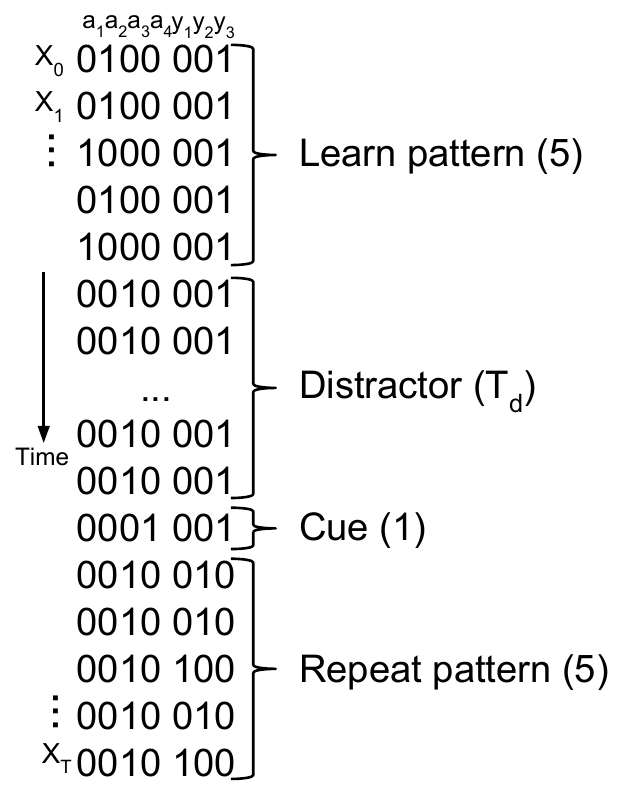}
\caption{Example data from the 5-bit task. The length of the sequence is $T$. The signals $a_1$, $a_2$, $a_3$ and $a_4$ are input-signals, while $y_1$, $y_2$ and $y_3$ are output-signals. In the first five time-steps the system learns the pattern. The system is then distracted for $T_d$ time-steps. After the cue-signal is set, the system is expected to reproduce the pattern that was learned. }
\label{fig:5bit}
\end{figure}

An example run with rule 90 is shown in fig. \ref{fig:rcca_single}. This visualisation gives valuable insights in how the reservoir behaves when parameters are changed, and makes it easier to understand the reservoir dynamics. 
Most natural systems come in the form of a temporal system (sequential), i.e., an input to the system depends on previous inputs. Classical feed-forward architectures are known to have issues with temporal tasks \cite{Goodfellow-et-al-2016-Book}.
In order to test the ReCA-system at a temporal task, the 5-bit task \cite{hochreiter1997long} is chosen in this paper. Such task has become a popular and widely used benchmark for reservoir computing, in particular because it tests the long-short-term memory of the system. 
An example data set from this task is presented in fig. \ref{fig:5bit}. The length of the sequence is given by $T$. $a_1, a_2, a_3$ and  $a_4$ are the input-signals, and $y_1, y_2$ and $y_3$ are the output-signals. At each time-step $t$ only one input-signal, and one output-signal, can have the value 1. The values of $a_1$ and $a_2$ at the first five time-steps give the pattern that the system shall learn. The next $T_d$ time-steps represent the distractor-period, where the system is distracted from the previous inputs. This is done by setting the value of $a_3$ to 1. After the disctractor period, the $a_4$ signal is fired which marks the cue-signal. The system is then asked to repeat the input-pattern on the outputs $y_1$ and $y_2$. The output $y_3$ is a waiting signal, which is supposed to be 1 right until the input-pattern is repeated. More details on the 5-bit memory task can be found in \cite{jaeger2012long}.

\subsection{Use of parallel CA-reservoirs in RC} \label{method.parallel}
\begin{figure}
\centering
    \includegraphics[width=3in]{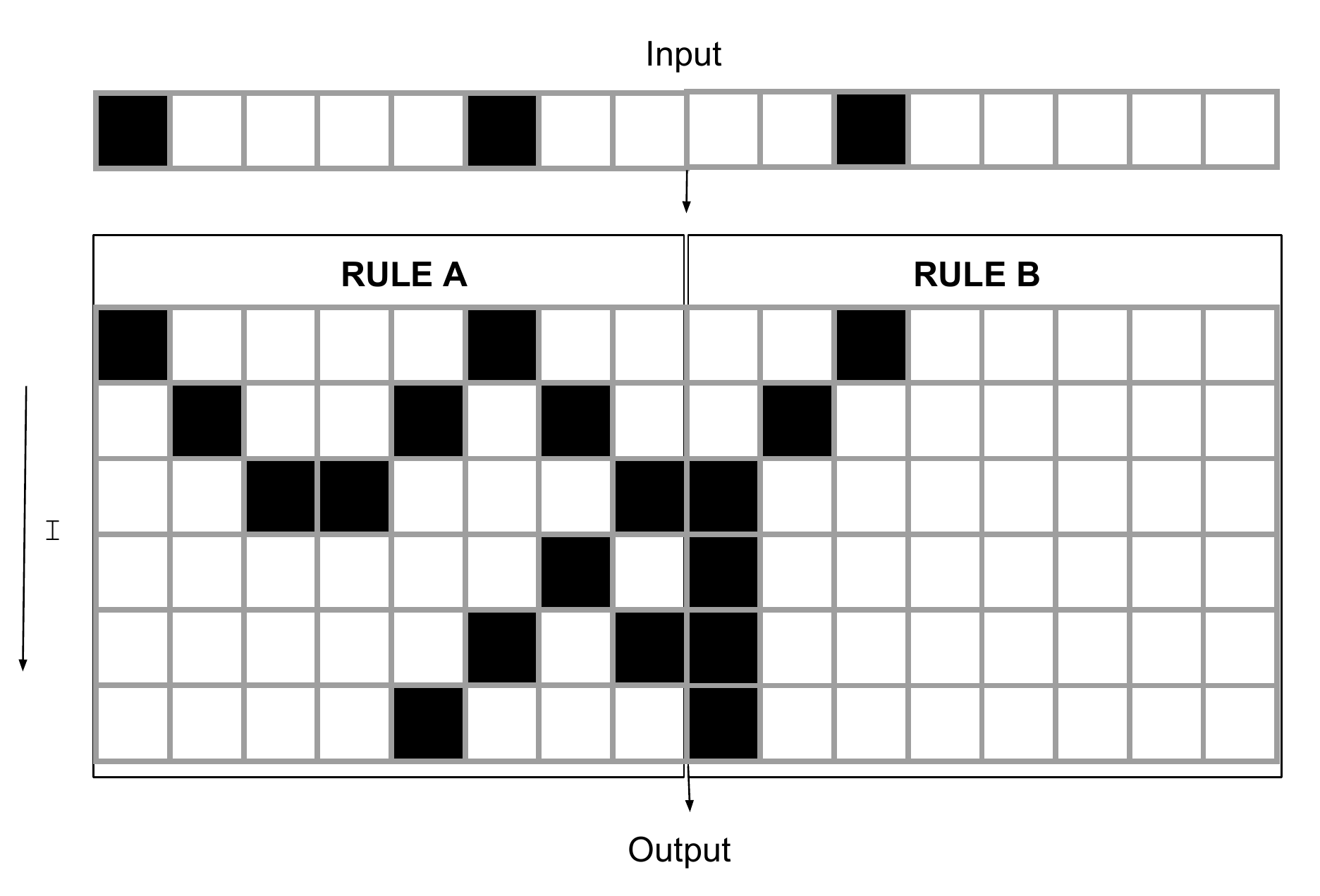}
\caption{Concept behind parallel CA reservoirs. Iterations flow downward. The rules are interacting at the middle boundaries and at the side boundaries, where the CA wraps around.}
\label{fig:nonUniCA}
\end{figure}

In this paper the use of parallel reservoirs is proposed. The concept is shown in fig. \ref{fig:nonUniCA}. At the  boundary conditions, i.e. the cell at the very end of the reservoir, the rule will treat the cell that lies within the other reservoir, as a cell in its own reservoir. This causes information/computation to flow between the reservoirs (loosely coupled). 

By having different rules in the reservoirs, one might be able to solve different aspects of the same problem, or even two problems at the same time. In \cite{bye2016master}, both the temporal parity and the temporal density task \cite{jaeger2012long} are investigated.  

\begin{figure}
\centering
    \includegraphics[height=3in]{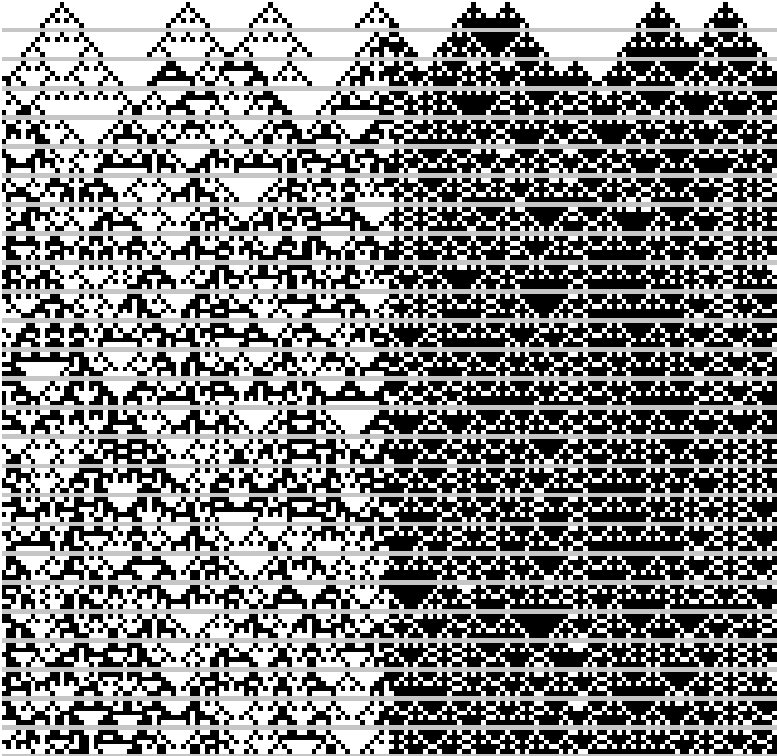}
\caption{Example run of the ReCA system with rule 90 on the left and 182 on the right. Information is allowed to flow between the reservoirs. The run is done with the parameters R=8, I=4 and C=5. The horizontal gray lines represent a time-step, in which the time-transition function is applied to every bit. Time flows downwards. The visualization is produced with the implemented system.}
\label{fig:rcca_parallel}
\end{figure}

Which rule is most suited for a task is still an open research question. The characteristics and classes described in section \ref{background.ca} are useful knowledge, however it does not precisely describe why some rules perform better than others on different tasks. In fig. \ref{fig:rcca_parallel} an example run of the parallel system is showed, with rule 90 on the left, and 182 on the right. This visualization gives useful insights on how the rules interact.

\subsection{Measuring computational complexity of a CA-reservoir}
The size of the reservoir is crucial for the success of the system. In this paper, the reservoir size is measured by $R*I*C$. As seen in section \ref{method.parallel}, the size of the reservoirs will remain the same both for the one-rule reservoirs and the two-rule reservoirs. This is crucial in order to be able to directly compare their performances.

\section{Results} \label{experiments.5bit}

\begin{table}
    \centering
    \caption{5-bit task parameters}
    \begin{tabular}{ l  l }
    Training set size   & 32 \\
    Testing set size    & 32 \\
    Distractor period   & 200 \\
    No. runs          & 120 \\
    \end{tabular}
    \label{tab:5bitparameters}
\end{table}

\begin{table}
    \centering
    \caption{CA reservoir parameter combinations}
    \begin{tabular}{ l  l }
    CA rules            & 60, 90, 102, 105, 150, 153, 165, 180, 195  \\
    I (iterations)      & 2, 4 \\
    R (random mappings) & 4, 8 \\
    C (size multiple)   & 10   \\
    \end{tabular}
    \label{tab:singleCAparam}
\end{table}

The parameters for the used 5-bit memory task can be seen in table \ref{tab:5bitparameters}.
The same parameters as in the single-reservoir system are used in the quasi-uniform CA reservoir system with a combination of two rules. 
The tested combinations of rules are shown in table \ref{tab:singleCAparam}.

\subsection{Results from the single ReCA-system}
The results from the single reservoir ReCA-system can be seen in table \ref{tab:resultsSingleCA}. The results in this paper are significantly better than what was reported in \cite{bye2016master}. We can however see a similar trend. Rules 102 and 105 were able to give promising results, while rule 180 was not very well suited for this task. An exception is rule 90 and 165, where the results in table \ref{tab:resultsSingleCA} show very high accuracy. In \cite{yilmaz2015connectionist} very promising results from rule 90 are also achieved.

\begin{table}
    \centering
    \caption{Single reservoir CA on 5-bit task. Successful runs with T=200}
    \begin{tabular}{ l  l  l  l  l}
    \textbf{Rule} & I=2, R=4  & I=2, R=8 &  I=4, R=4 &  I=4, R=8 \\
        60&		    25.8\%&		53.3\%&		76.7\%&		95.0\%	\\
        90&		    100.0\%&	100.0\%&	97.5\%&		100.0\%\\
        102&		30.8\%&		63.3\%&		71.7\%&		96.7\%	\\
        105&		95.8\%&		99.2\%&		99.2\%&		100.0\%\\
        150&		96.7\%&		100.0\%&	100.0\%&	100.0\%\\
        153&		26.7\%&		55.0\%&		80.0\%&		95.0\%	\\
        165&		100.0\%&	100.0\%&	100.0\%&	100.0\%\\
        180&		9.2\%&		38.3\%&		0.8\%&		1.7\%	\\
        195&		39.2\%&		61.7\%&		79.2\%&		95.8\%	\\
    \end{tabular}
    \label{tab:resultsSingleCA}
\end{table}

\subsection{Results from the parallel (non-uniform) ReCA-system} \label{results.parallel}
Results can be seen in table \ref{tab:resultsNonUniformCA}. It can be observed that combination of rules that were performing well in table \ref{tab:resultsSingleCA} seem to give good results when combined. However, some combination of rules, e.g., 60 and 102, 153 and 195, gave worse results than the rules by themselves. We can observe the same tendencies as in the single-runs; higher R and I generally yields better results. 

\begin{table}
    \centering
    \caption{Parallel CA on 5-bit task. Successful runs with T=200}
    \begin{tabular}{ l  l  l  l  l}
    \textbf{Rule} &     I=2, R=4  & I=2, R=8 &  I=4, R=4 &  I=4, R=8 \\
        60 and 90&      87.5\%&		100.0\%&	96.9\%&		100.0\%\\
        60 and 102& 	0.0\%&		0.0\%&		0.0\%&		0.0\%\\
        60 and 105&	    81.2\%&		100.0\%&	96.9\%&		100.0\%\\
        60 and 150&	    71.9\%&		100.0\%&	96.9\%&		100.0\%\\
        60 and 153&	    0.0\%&		0.0\%&		0.0\%&		0.0\%\\
        60 and 165&	    87.5\%&		93.8\%&		96.9\%&		96.9\%\\
        60 and 180&	    43.8\%&		53.1\%&		90.6\%&		84.4\%\\
        60 and 195&	    0.0\%&		0.0\%&		0.0\%&		0.0\%\\
        90 and 102&	    90.6\%&		100.0\%&	100.0\%&	96.9\%\\
        90 and 105&	    100.0\%&	100.0\%&	100.0\%&	100.0\%\\
        90 and 150&	    100.0\%&	100.0\%&	100.0\%&	100.0\%\\
        90 and 153&	    93.8\%&		96.9\%&		96.9\%&		100.0\%\\
        90 and 165&	    90.6\%&		100.0\%&	100.0\%&	100.0\%\\
        90 and 180&	    90.6\%&		100.0\%&	100.0\%&	100.0\%\\
        90 and 195&	    87.5\%&		96.9\%&		100.0\%&	100.0\%\\
        102 and 105&	78.1\%&		100.0\%&	96.9\%&		100.0\%\\
        102 and 150&	81.2\%&		100.0\%&	96.9\%&		100.0\%\\
        102 and 153&	0.0\%&		0.0\%&		0.0\%&		3.1\%\\
        102 and 165&	93.8\%&		100.0\%&	100.0\%&	100.0\%\\
        102 and 180&	0.0\%&		40.6\%&		3.1\%&		6.2\%\\
        102 and 195&	0.0\%&		0.0\%&		0.0\%&		3.1\%\\
        105 and 150&	93.8\%&		100.0\%&	100.0\%&	100.0\%\\
        105 and 153&	75.0\%&		93.8\%&		93.8\%&		100.0\%\\
        105 and 165&	96.9\%&		100.0\%&	100.0\%&	100.0\%\\
        105 and 180&	93.8\%&		100.0\%&	100.0\%&	100.0\%\\
        105 and 195&	65.6\%&		93.8\%&		96.9\%&		100.0\%\\
        150 and 153&	87.5\%&		100.0\%&	96.9\%&		100.0\%\\
        150 and 165&	100.0\%&	100.0\%&	100.0\%&	100.0\%\\
        150 and 180&	81.2\%&		100.0\%&	100.0\%&	100.0\%\\
        150 and 195&	78.1\%&		96.9\%&		100.0\%&	100.0\%\\
        153 and 165&	81.2\%&		100.0\%&	100.0\%&	100.0\%\\
        153 and 180&	3.1\%&		46.9\%&		0.0\%&		0.0\%\\
        153 and 195&	0.0\%&		0.0\%&		0.0\%&		0.0\%\\
        165 and 180&	96.9\%&		96.9\%&		100.0\%&	100.0\%\\
        165 and 195&	87.5\%&		100.0\%&	100.0\%&	100.0\%\\
        180 and 195&	40.6\%&		87.5\%&		93.8\%&		96.9\%\\
    \end{tabular}
    \label{tab:resultsNonUniformCA}
\end{table}

\section{Analysis}

\subsection{Single reservoir ReCA-system}

The complexity of the reservoir is a useful metric when comparing different approaches. If we examine rule 90, we can observe that it achieves 100\% success rate at $I=4$, $R=8$ and $C=10$. The size of the reservoir is ${4*8*10 = 320}$ at this configuration. Note that even though lower values of $R$ and $I$ also give 100\%, at $R=4$ and $I=4$ the success is  97.5\%, yet again 100\% at $I=4$ and $R=8$. \cite{yilmaz2015connectionist} reported a 100\% success-rate on the same task with $R=32$ and $I=16$. The C-parameter was set to 1. As such, the size of the reservoir is ${32*16*1 = 512}$ (feed-forward architecture). 

\cite{yilmaz2015connectionist} also presented results on the 5-bit task using the recurrent architecture. 100\% success-rate was achieved with $I=32$ and $R=45$. This yields a reservoir size of ${32*45 = 1440}$. Those results were intended to study the relationship between the distractor period of the 5-bit task, and the $R$ number of random mappings. The $I$ was kept fixed at 32 during this experiment. Even if the motivation for the experiments were different, the comparison of results gives insight that the reservoir size itself may not be the only factor that determines the performance of the ReCA system.  

\subsection{Parallel reservoir (non-uniform) ReCA-system}

Why are some combinations better than others? As observed in section \ref{results.parallel}, rules that are paired with others rules that perform well on their own, also perform well together. The combination of rule 90 and rule 165 is observed to be very successful. As described in \cite{wolfArule90}, rule 165 is the complement of rule 90. If we observe the single-CA results in table \ref{tab:resultsSingleCA} we can see that rule 90 and 165 perform very similarly. 

\begin{figure}
\centering
    \includegraphics[height=3in]{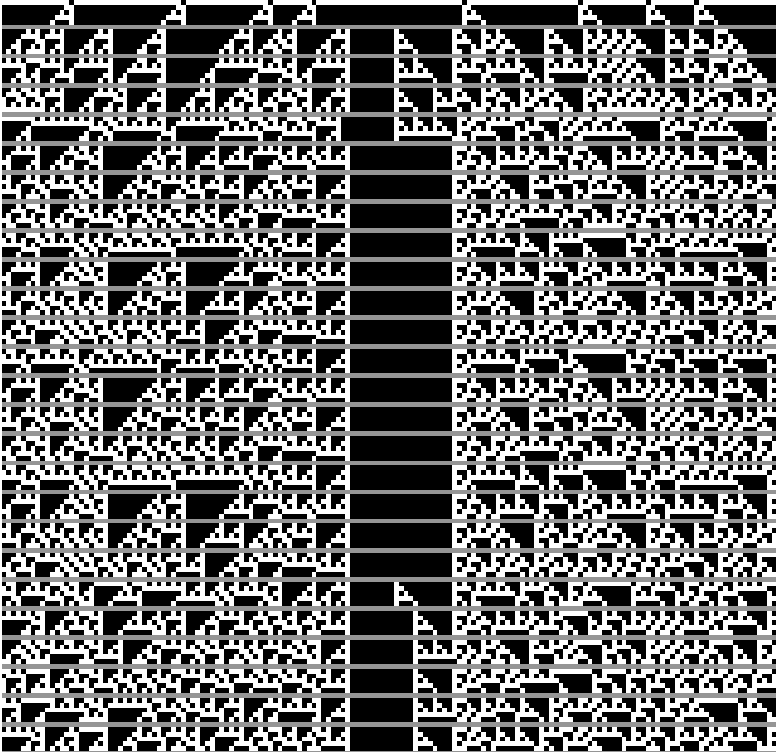}
\caption{Example run of the ReCA system with rule 153 and 195. The run is done with the parameters R=8, I=4 and C=5. The horizontal gray lines represent a time-step, in which the time-transition function is applied to every bit. Time flows downwards. The visualization is produced with the implemented system.}
\label{fig:disc.bad}
\end{figure}

Examining one of the worst-performing rule-combinations of the experiments, i.e., rule 153 and rule 195, we get some useful insight as seen in fig. \ref{fig:disc.bad}. Here it is possible to notice that the interaction of rules creates a "black" region in the middle (between the rules), thereby effectively reducing the size of the reservoir. As described in \cite{wolfArule60}, rule 153 and 192 are the mirrored complements. 

Rule 105 is an interesting rule to be combined with others. As described in \cite{wolfram2002new}, the rule does not have any compliments or any mirrored compliments. Nevertheless, as seen in table \ref{results.parallel}, it performs well in combination with most other rules.

\section{Conclusion}
A framework for using cellular automata in reservoir computing has been implemented, which makes use of uniform CA and quasi-uniform CA. Relationship between reservoir size and performances of the system are presented. The implemented configuration using parallel CA reservoir is tested in this paper for the first time (to the best of the authors' knowledge). Results have shown that some CA rules work better in combination than other. Good combinations tend to have some relation, e.g. being complementary. Rules that are mirrored compliments do not work well together, because they effectively reduce the size of the reservoir. The concept is still very novel, and a lot of research is left to be done, both regarding the use of non-uniform CA reservoir, as well as ReCA-systems in general. 

As previously discussed, finding the best combination of rules is not trivial. If we only consider the usage of two distinct rules, the rule space grows from only $256$ single-reservoir options to ${\frac{256!}{2!*254!}=32640}$ different combinations. Matching two rules that perform well together can be quite a challenge. By investigating the characteristics of the rules, e.g., with lambda-parameter \cite{langton1990computation}, Lyapunov exponent \cite{legenstein2007edge} or other metrics, it may be possible to pinpoint promising rules. Ideally, the usage of more than two different rules could prove a powerful tool. The rule space would then grow even larger, and an exhaustive search would be infeasible. However, one possibility would be to use evolutionary algorithms to search for suitable rules. Adding more and more rules would bring the reservoir closer to a true non-uniform CA. 

In \cite{jaeger2012long} a wide range of different tasks is presented. In this paper only one (5-bit task) is used as a benchmark. By combining different rules' computational power, one could design a reservoir that performs well on a variety of tasks.

\appendices


\ifCLASSOPTIONcaptionsoff
  \newpage
\fi



\bibliographystyle{plain}
\bibliography{references}

\end{document}